\documentclass[11pt,prd,preprint,superscriptaddress,amsmath,amssymb,nofootinbib]{revtex4}

\usepackage{graphicx}
\usepackage{dcolumn}
\usepackage{bm}
\usepackage{amssymb}
\usepackage{amsmath}
\usepackage{epsfig}    
\usepackage{color}
\usepackage{slashed}
\usepackage{hhline}
\usepackage{ulem}

\def\be{\begin{equation}}
\def\ee{\end{equation}}
\newcommand{\bea}{\begin{eqnarray}}
\newcommand{\eea}{\end{eqnarray}}
\newcommand{\nn}{\nonumber}

\begin{document}


\title{Lepton seesaw model in a modular $A_4$ symmetry}


\author{Takaaki Nomura}
\email{nomura@scu.edu.cn}
\affiliation{College of Physics, Sichuan University, Chengdu 610065, China}

\author{Hiroshi Okada}
\email{hiroshi3okada@htu.edu.cn}
\affiliation{Department of Physics, Henan Normal University, Xinxiang 453007, China}

\date{\today}

\begin{abstract}
{
We propose a lepton seesaw model to get the same mass-origin of charged-leptons and neutrinos introducing a modular $A_4$ symmetry. 
In this scenario, the mass matrix of charged-leptons is induced via seesaw-like mechanism while the mass matrix of neutrino is realized via an inverse seesaw mechanism with help of vector-like fermions and an isospin singlet and triplet scalar fields with nonzero vacuum expectation values. In analyzing the model we concentrate on fixed points of modulus $\tau$ that is favored by a flux compactification of Type IIB string theory.
We search for allowed region at nearby these fixed points and find good solutions at nearby $\tau=\omega$ with several good predictions.  
}
 %
 \end{abstract}
\maketitle
\newpage

\section{Introduction}
Even though charged-leptons and the neutrinos are in the same component in the left-handed $SU(2)$ doublet of the standard model (SM), their masses are often discussed with different manners. 
This is because the masses of neutrinos are zero in the SM and we need to extend the SM so that we have (tiny but) {non-zero} neutrino masses.
One of the simplest ways is to introduce heavy right-handed neutrinos that is analogy of the other three sectors, and the small neutrino masses can simply be realized via seesaw mechanism~\cite{Yanagida:1979gs, Minkowski:1977sc,Mohapatra:1979ia,Zee:1980ai}. But in this scenario, there is no relation between the mass matrix of neutrinos and charged-leptons.
{Thus it is interesting if we find a unified description of generating masses of them.}

In this paper, we study a possibility of the same origin for the charged-leptons and the neutrino mass matrices~\cite{Lee:2021gnw}, which is called "lepton seesaw",  introducing several exotic fermions (vector-like fermions) and scalar fields (an isospin triplet scalar of hyper-charge 1 and singlet scalar of hyper-charge zero with nonzero vacuum expectation values). In order to realize our model, we impose a modular flavor $A_4$ group that has a promising symmetry to get our desired terms as well as restricting some forms of mass matrices.
Since 2017~\cite{Feruglio:2017spp}, this symmetry, {especially modular $A_4$}, has widely been applied to various phenomenology in flavor physics~\cite{Criado:2018thu, Kobayashi:2018scp, Okada:2018yrn, Nomura:2019jxj, Okada:2019uoy, deAnda:2018ecu, Novichkov:2018yse, Nomura:2019yft, Okada:2019mjf,Ding:2019zxk, Nomura:2019lnr,Kobayashi:2019xvz,Asaka:2019vev,Zhang:2019ngf, Gui-JunDing:2019wap,Kobayashi:2019gtp,Nomura:2019xsb, Wang:2019xbo,Okada:2020dmb,Okada:2020rjb, Behera:2020lpd, Behera:2020sfe, Nomura:2020opk, Nomura:2020cog, Asaka:2020tmo, Okada:2020ukr, Nagao:2020snm, Okada:2020brs,Kang:2022psa, Ding:2024fsf, Ding:2023htn, Nomura:2023usj, Kobayashi:2023qzt, Petcov:2024vph, Kobayashi:2023zzc, Nomura:2024ghc, Qu:2024rns}.
{Furthermore, a framework of non-holomorphic modular symmetry was recently developed~\cite{Qu:2024rns} which helps to construct simpler models without supersymmetry.}
{In the model of this work, the {non-holomorphic} modular $A_4$ symmetry plays two important roles; (1) Forbidding unnecessary terms to realize lepton seesaw, (2) Providing predictions constraining structure of Yukawa couplings.}
{After constructing the model we perform numerical analysis to find some predictions in neutrino sector imposing experimental constraints.}
In our chi square numerical analysis, we concentrate on some specific regions at nearby fixed points of modulus $\tau$ that is a new degree of freedom via modular symmetry. These fixed points are favored by a flux compactification of Type IIB string theory~\cite{Ishiguro:2020tmo}.
We find allowed regions at nearby $\tau=\omega$ that preserves a remnant $Z_3$ symmetry, and obtain several predictions such as phases and masses in neutrino sector.

This paper is organized as follows.
In Sec. \ref{sec:II}, 
we review our setup of the modular $A_4$ assignments introducing new particles and explain how to realize our scenario by showing the mass generation mechanisms for charged-fermion sector and the neutral fermion sector. 
In Sec. \ref{sec:III}, we demonstrate chi-square numerical analyses concentrating on fixed points and show what kinds of predictions we obtain.
Finally, we summarize and conclude in Sec. \ref{sec:IV}.

\section{Model setup}
\label{sec:II}

\begin{table}[t!]
\begin{tabular}{|c||c|c|c|c|c|c|c|}\hline\hline  
& ~$L_L$~ & ~$ \overline{\ell_R}$~& ~$ \overline{L'_R}$~& ~$ {L'_L}$~ & ~$ H$ ~&~ {$\Delta$}~~&~ {$\varphi$}~  \\\hline\hline 
$SU(2)_L$   & $\bm{2}$  & $\bm{1}$  & $\bm{2}$ & $\bm{2}$ & $\bm{2}$  & $\bm{3}$    & $\bm{1}$  \\\hline 
$U(1)_Y$    & $-\frac12$  & $1$  & $\frac12$ & $-\frac12$  & $\frac12$ & $1$& $0$    \\\hline
$A_4$   & $\bm{3}$  & $ \{ \bar{\bm{1}} \}$ & $\{ \bar{\bm{1}} \}$ & $\{ \bm{1} \} $  & $\bm{1}$ & $\bm{1}$ & $\bm{1}$         \\\hline 
$-k_I$    & ${-1/2}$  & $-2$& $-2$ & $+2$ & $0$ & ${0}$  & ${1/2}$     \\\hline
\end{tabular}
\caption{Charge assignments of the fermions and bosons
under $SU(2)_L\otimes U(1)_Y \otimes A_4$ where $-k_I$ is the number of modular weight. Here, {$\{ \bm{1} \} =\{1, 1', 1''\}$} indicates assignment of $A_4$ singlets.}\label{tab:1}
\end{table}
Here, we review our model {based on the non-holomorphic modular $A_4$ symmetry}.
At first, we introduce three isospin doublet {vector-like leptons} $L'(\equiv[N,E]^T$) with hypercharge $Y=-1/2$, an isospin triplet scalar field $\Delta$ with $Y=1$,
and an isospin singlet scalar field $\varphi(\equiv (v'+r)/\sqrt2)$  with $U(1)_Y= 0$, in addition to the SM fields. The SM Higgs is denoted by $H$ and its vacuum expectation value (VEV) is defined by $\langle H\rangle=[0,v/\sqrt2]^T$.
{We also require $\varphi$ and $\Delta$ to develop VEVs denoted by $v'$ and $v_\Delta$ respectively.}
The SM right-handed leptons $\ell_R$, and  $L'$ are assigned to be {$\{ \bm{1} \} =\{1, 1', 1''\}$} under the $A_4$ symmetry, while the left-handed SM leptons $L_L\equiv[\nu_L,\ell_L]^T$ are assigned to be triplet under the $A_4$ symmetry. 
\footnote{{
{Although we adopt the non-holomorphic modular $A_4$ symmetry in this work, it is possible to obtain the same structure using holomorphic case.}
When one 
{needs to} work on a supersymmetric theory, main discussions below are totally maintained by changing assignments for modular weight into $+3$ for  $L_L$, $\Delta$ for $-4$, and $-3$ for $\varphi$.
}}
All the scalar fields are trivial singlet under the $A_4$ symmetry. In addition, we impose the modular weight for each field; {$-1/2$ for $L_L$, $-2$ for $ \overline{\ell_R}$, $-2$ for $ \overline{L'_R}$, $+2$ for ${L'_L}$, $0$ for $H$, $0$ for $\Delta$, and $1/2$ for $\varphi$.}
All fields and their charge assignments are summarized in TABLE~\ref{tab:1}.
These assignments forbid undesired mass terms at tree-level such as $\overline{\ell_R} {\tilde H} L_L$, $\overline{L^C_L}(i\tau_2) \Delta L_L$ and $\overline{L^C_L}(i\tau_2) \Delta L'_L$
{due to non-even number of modular weights.}

Under these symmetries, the renormalizable Lagrangian is given by  
\begin{align}
-{\cal L}_\ell =
 [ \overline{\ell_R} {H^\dagger} L'_L] 
+
[Y^{(2)}_3 \overline{L'_R} L_L\varphi ] 
+
{
[ Y^{(-4)}_1\overline{L'^C_L}(i\tau_2) \Delta L'_L] 
}
+
{
[Y^{(4)}_{1,1'} \overline{L'_R}(i\tau_2) \Delta^\dag L'^C_R] 
}
+
[M_{L'}  \overline{L'_R} L'_L]
+{\rm h.c.},
\label{eq:lpy}
\end{align}
where $\tau_2$ is the second Pauli matrix, and $[\cdots]$ represents a trivial $A_4$ singlet ${\bm 1}$ abbreviating the free parameters that are explicitly shown later. $Y^{(2)}_3\equiv [y_1,y_2,y_3]^T$ is modular coupling that is uniquely fixed when modulus $\tau$ is determined~\cite{Feruglio:2017spp}.
The Higgs potential is given by
\begin{align}
{\cal V} &=
-\mu_H^2 |H|^2 + \mu_\Delta^2 {\rm Tr}[\Delta^\dag\Delta] - \mu_\varphi^2|\varphi|^2
+\mu (H^T(i\tau_2) \Delta^\dag H +{\rm h.c.})\nn\\
&+\lambda_H |H|^4 +\lambda_\Delta  {\rm Tr}[|\Delta^\dag\Delta|^2]+\lambda'_\Delta  {\rm det}[\Delta^\dag\Delta]
+ \lambda_\varphi |\varphi|^4 { + ( \tilde{\lambda}_\varphi Y_1^{(-2)} \varphi^4 + h.c.) }
\nn \\ 
& +\lambda_{H\Delta} |H|^2  {\rm Tr}[\Delta^\dag\Delta]  
 +\lambda'_{H\Delta} \sum_i^3 (H^\dag\tau_i H)  {\rm Tr}[\Delta^\dag \tau^i\Delta] {+ \lambda_{H \varphi} |H|^2 |\varphi|^2 }
+\lambda_{\Delta\varphi} {\rm Tr}[\Delta^\dag\Delta] |\varphi|^2,
\label{eq:lpy}
\end{align}
where $\tau_i\ (i=1,2,3)$ are Pauli matrices and some of free parameters include modular forms; e.g.,
$\mu_\varphi \equiv \mu'_\varphi (i\tau-i\tau^*)^{1/2}$.
The scalar potential is similar to that of Higgs triplet model with a little modification due to one scalar singlet extension. 
{
The scalar fields can be written by 
\begin{equation}
H = 
\begin{pmatrix}
G^+ \\
\frac{1}{\sqrt{2}} (v + h + i G) 
\end{pmatrix}, \quad 
\Delta = \begin{pmatrix} \frac{\delta^+}{\sqrt{2}} & \delta^{++} \\ \frac{1}{\sqrt{2}} (v_\Delta + \delta^0 + i \eta) & -\frac{\delta^+}{\sqrt{2}}   \end{pmatrix}, \quad
\varphi = \frac{1}{\sqrt{2}} (v' + \phi + i \chi),
\end{equation}
where $G^\pm$ and $G$ are Nambu-Goldstone bosons absorbed by the massive SM gauge fields $W^\pm$ and $Z$.
The VEVs of the scalar fields are obtained by solving the conditions
\begin{align}
\frac{\partial V}{\partial v}\Bigg|_0 = \frac{\partial V}{\partial v'}\Bigg|_0 = \frac{\partial V}{\partial v_\Delta}\Bigg|_0 =0, 
\end{align}
with $|_0$ indicating all the component fields to be zero. 
The stable vacuum can be obtained by assuming $\mu^2_{\{H, \Delta, \varphi\}} >0$, $\mu>0$, $\lambda_{\{H, \varphi\}}>0$ where we also assume 
$\{\tilde \lambda_{\varphi}, \lambda_{H \varphi}\} \to 0$ for simplicity.
The VEVs are then obtained such that
\begin{equation}
v \simeq \sqrt{\frac{\mu^2_H}{\lambda_H}}, \quad v' \simeq \sqrt{\frac{\mu^2_\varphi}{\lambda_\varphi}}, \quad v_\Delta \simeq \frac{\mu v^2}{\sqrt{2} \mu_\Delta^2}. 
\end{equation}
The triplet VEV can be small taking $v^2 \ll \mu^2_\Delta$ by type-II seesaw mechanism.
The masses of scalar fields are also simply given by 
\begin{equation}
m_h \simeq \sqrt{\lambda_H} v, \quad m_{\{\phi, \chi \}} \simeq \sqrt{\lambda_\varphi} v', \quad m_{\{\delta^0,\eta,\delta^+, \delta^{++}\}} \simeq \mu_\Delta.
\end{equation}
In our analysis below the extra scalar masses are not important and we just choose them to be larger than TeV scale in order to satisfy current collider constraints.
}

\subsection{Charged-lepton mass matrix}
The mass matrix of charged-lepton originates from the following terms after spontaneous electroweak symmetry breaking;
\begin{align}
& [ \overline{\ell_R} { H^\dagger} L'_L] 
+
[Y^{(2)}_3 \overline{L'_R} L_L\varphi ] 
+
[M_{L'}  \overline{L'_R} L'_L] \nonumber \\
& \equiv
{ a_e} \overline{e_R}  {H^\dagger} L'_{L_1} + { b_e} \overline{\mu_R}  { H^\dagger} L'_{L_2} + { c_e} \overline{\tau_R}  { H^\dagger} L'_{L_3} \nonumber \\
& +
f_1 \overline{L'_{R_1}} [y_1 L_{L_1}+y_2 L_{L_3}+y_3 L_{L_2} ]\varphi
{ +
f_2  \overline{L'_{R_2}} [y_3 L_{L_3}+y_1 L_{L_2}+y_2 L_{L_1} ]\varphi }\nonumber \\
& { +
f_3 \overline{L'_{R_3}} [y_2 L_{L_2}+y_1 L_{L_3}+y_3 L_{L_1} ]\varphi
}
 +
M_1 \overline{L'_{R_1}}  L'_{L_1}
+
M_2 \overline{L'_{R_2}}  L'_{L_2}
+
M_3 \overline{L'_{R_3}}  L'_{L_3}\\
&\supset 
m \overline{\ell_R}   E_L
+m'  \overline{E_R} \ell_L
+M_{L'}  \overline{E_R} E_L,
\label{eq:cgd-lep}
\end{align}
where each form of $m$, $m'$, and $M_{L'}$ is given by
\begin{align}
m = \frac{v}{\sqrt2}
 \left(\begin{array}{ccc} 
a_e & 0 & 0 \\
0 & b_e& 0\\
0& 0 & c_e  \end{array} \right),\
m' = \frac{v'}{\sqrt2}
 \left(\begin{array}{ccc} 
f_1 & 0 & 0 \\
0 & f_2 & 0\\
0& 0 & f_3  \end{array} \right)
 \left(\begin{array}{ccc} 
y_1 & y_3 & y_2 \\
y_2 & y_1 & y_3\\
y_3& y_2 & y_1  \end{array} \right),\
M_{L'}=  
 \left(\begin{array}{ccc} 
M_1 & 0 & 0 \\
0 & M_2 & 0\\
0& 0 & M_3  \end{array} \right),
\label{eq:cgd-lep2}
\end{align}
where $\{a_e, b_e, c_e \}$, $\{f_1,f_2,f_3\}$, and $\{M_1,M_2,M_3\}$ can be real parameters without loss of generality.
Then, in basis of $[\ell_L,E_L]^T$, the mass matrix of charged-lepton sector is given by
\begin{align}
{\cal M}_e= \left(\begin{array}{cc} 
0 & m \\
m' &M_{L'}
  \end{array} \right) .
\label{eq:cgd-mtrx}
\end{align}
When  we assume relations $m,m' \ll M_{L'}$ among mass matrices,~\footnote{{It is reasonable assumption since $M_{L'}$ is mass parameter in original Lagrangian and would be large. However these hierarchies cannot be controlled by any flavor symmetry and one might reply on the other mechanism such as radiatively induced scenarios, e.g. ref.~\cite{Das:2017ski}. Here, it is beyond our scope.}}
we can treat the charged-lepton mass matrix as a seesaw-like as follows.
First of all, ${\cal M}_e$ is diagonalized by
\begin{align}
\left(\begin{array}{cc} 
m_e & 0 \\
0 &M_{e}
  \end{array} \right)= 
  O^\dag_R{\cal M}_e O_L.
\end{align}
Due to the mass hierarchies among $m$, $m'$ and $M_{L'}$, we write the mixing matrix and mass matrices as follows:
\begin{align}
& O_{L}\simeq
\left(\begin{array}{cc} 
1 & -\theta_{L}\\
\theta_{L} & 1
  \end{array} \right),\quad
 \theta_{L}\approx -M_{L'}^{-1} m',\nn\\
& m_e\approx m M_{L'}^{-1} m',\quad M_e\approx M_{L'}.
\end{align}
Since $M_{L'}$ is diagonal, we do not need to diagonalize it more.
$m_e$ is diagonalized by $D_e=V_{e_R}^\dag m_e V_{e_L}$. 
Therefore, we obtain $|D_e|^2=V^\dag_{e_L} m^\dag_e m_e V_{e_L}$ where $D_e=$
diag.$(m_e,m_\mu,m_\tau)$.
Since $m_e$ is given by
\begin{align}
 \frac{v v'}{2}
 \left(\begin{array}{ccc} \frac{a_e f_1}{M_1} & 0 & 0 \\
0 & \frac{b_e f_2}{M_2} & 0\\
0& 0 & \frac{c_e f_3}{M_3}  \end{array} \right)
 \left(\begin{array}{ccc} 
y_1 & y_3 & y_2 \\
y_2 & y_1 & y_3\\
y_3& y_2 & y_1  \end{array} \right),
\label{eq:cgd-lep2}
\end{align}
$\{\frac{a_e f_1}{M_1}, \frac{b_e f_2}{M_2},  \frac{c_e f_3}{M_3} \}$ are fixed to fit the three observed charged-lepton masses by the following relations:
\begin{align}
&{\rm Tr}[m^\dag_e m_e] = |m_e|^2 + |m_\mu|^2 + |m_\tau|^2,\quad
 {\rm Det}[m^\dag_e m_e] = |m_e|^2  |m_\mu|^2  |m_\tau|^2,\nn\\
&({\rm Tr}[m^\dag_e m_e)^2 -{\rm Tr}[(m^\dag_e m_e)^2] =2( |m_e|^2  |m_\mu|^2 + |m_\mu|^2  |m_\tau|^2+ |m_e|^2  |m_\tau|^2 ).\label{eq:l-cond}
\end{align}

\subsection{Neutral fermion mass matrix}
The mass matrix of neutral fermions comes from the following terms
after spontaneous electroweak symmetry breaking;
\begin{align}
&[\overline{L'^C_L}(i\tau_2) \Delta L'_L] 
+
[\overline{L'_R}(i\tau_2) \Delta^\dag L'^C_R] 
+
[Y^{(2)}_3 \overline{L'_R} L_L\varphi ] 
+
[M_{L'}  \overline{L'_R} L'_L]
 \nn\\
&\supset
\delta_N \overline{N_L^C}   N_L
+\delta_{\bar N} \overline {N_R}   N_R^C
+m'  \overline{N_R} \nu_L
+M_{L'}  \overline{N_R} N_L,
\label{eq:cgd-neut}
\end{align}
where $\delta_N$ is given by
\begin{align}
\delta_N = 
\frac{v_{\Delta}}{\sqrt2}
 \left(\begin{array}{ccc} 
y_\Delta & 0 & 0 \\
0& 0 & y'_\Delta  \\
0 & y'_\Delta & 0 \end{array} \right)
= 
\frac{v_{\Delta} y_\Delta}{\sqrt2}
 \left(\begin{array}{ccc} 
1 & 0 & 0 \\
0& 0 & \tilde y'_\Delta  \\
0 &  \tilde y'_\Delta & 0 \end{array} \right)
\equiv \frac{v_{\Delta} y_\Delta}{\sqrt2} \tilde\delta_N
 ,
\label{eq:cgd-lep2}
\end{align}
and $\delta_{\bar N} $ is three by three symmetric matrix with complex values.
Note here that $\delta_{\bar N} $ does not appear in the neutrino mass matrix as can be seen later.
Thus, we will not explicitly write it down. 
The resultant neutral fermion mass matrix in basis of $[\nu_L,N_L,N^C_R]^T$ is found as follows:
\begin{align}
 \left(\begin{array}{ccc} 
0 & 0 & m'^T \\
0& \delta_N & M_{L'}^T  \\
m' &  M_{L'} & \delta_{\bar N} \end{array} \right).
\label{eq:cgd-lep2}
\end{align}
When we impose the following mass hierarchies $\delta_{ N} \ll m' \ll M_{L'}$, one finds the neutrino mass matrix via inverse seesaw as follows:
\begin{align}
m_\nu = (M^{-1}_{L'} m')^T \delta_N (M^{-1}_{L'} m').
\label{eq:cgd-lep2}
\end{align}
As we can see, $M_{L'},\ m'$ also contributes to the charged-lepton mass matrix.
{Here we comment on our assumption regarding hierarchy of mass parameters; $\delta_{ N} \ll  m' \ll M_{L'}$. 
This can be appropriate assumption as follows. Firstly $\delta_{ N}$ is generated via $v_\Delta$ that is restricted to be smaller than a few GeV to satisfy the constraint on the $\rho$ parameter~\cite{ParticleDataGroup:2020ssz}, and it can be naturally small due to type-II seesaw mechanism.
The mass scale of $m'$ is determined by VEV $v'$ which would not be far from the electroweak scale where we would choose TeV scale.
Finally, scale of $M_{L'}$ is free and can be much larger than the electroweak scale. 
Note also that large $M_{L'}$ scale ($\mathcal{O}(100)$ TeV) is motivated to avoid constraints from LFV decays of charged leptons as discussed below.
}

{The mass eigenvalues for the active neutrinos $D_\nu = \{D_{\nu_1}, D_{\nu_2}, D_{\nu_3} \}$ are obtained by diagonalizing the mass matrix as $D_\nu \equiv V_\nu^T m_\nu V_\nu$.
Therefore, we obtain $V_\nu^\dag m_\nu^\dag m_\nu V_\nu ={\rm diag.}(|D_{\nu_1}|^2,|D_{\nu_2}|^2,|D_{\nu_3}|^2)$ where $D_{\nu_{1,2,3}}$ are the three neutrino mass eigenvalues.
{
We adopt the standard parametrization of the Pontecorvo-Maki-Nakagawa-Sakata (PMNS) mixing matrix {$U$}
defined by $V_{e_L}^\dag V_\nu$
where the Majorana phase is defined by $[1,e^{i\alpha_{21}/2},e^{i\alpha_{31}/2}]$~\cite{Okada:2019uoy}. 
Then, the Majorana phases are found as 
\begin{align}
\cos \left( \frac{\alpha_{21}}{2} \right) = \frac{\text{Re}[U^*_{e1} U_{e2}] }{ c_{12} s_{12} c_{13}^2 }, \
\cos \left( \frac{\alpha_{31}}{2} - \delta_{CP} \right) = \frac{\text{Re}[U^*_{e1} U_{e3}] }{ c_{12} s_{13} c_{13}} , \\
 \sin \left( \frac{\alpha_{21}}{2} \right) = \frac{\text{Im}[U^*_{e1} U_{e2}] }{ c_{12} s_{12} c_{13}^2 }, \
\sin \left( \frac{\alpha_{31}}{2} - \delta_{CP} \right) =\frac{ \text{Im}[U^*_{e1} U_{e3}] }{ c_{12} s_{13} c_{13} },
\end{align}
where $\alpha_{21}/2,\ \frac{\alpha_{31}}{2} - \delta_{CP}$
are subtracted from $\pi$, when $\cos(\alpha_{21}/2)$ and $\cos \left( \frac{\alpha_{31}}{2} - \delta_{CP} \right)$ are negative.
Similar to Majorana phases, $\delta_{CP}$ is given by
\begin{align}
\cos\delta_{CP} &=\frac{-1}{2c_{12}s_{12}c_{23}s_{23}s_{13}} (|U_{31}|^2 - s_{12}^2 s_{23}^2-c_{12}^2c_{23}^2 s_{13}^2), \\
\sin\delta_{CP} &=\frac{1}{s_{23}c_{23}s_{12}c_{12}s_{13}c_{13}^2} {\rm Im}[U_{11}  U_{22} U_{12}^* U_{21}^* ]. \
\end{align}
}
For convenience, we define dimensionless neutrino mass matrix $m_\nu \equiv\kappa \tilde m_\nu$
where $\kappa\equiv\frac{v_{\Delta} y_\Delta}{\sqrt2}$
is a flavor independent  mass dimensional parameter. 
Therefore, $\kappa$ can be rewritten in terms of rescaled neutrino mass eigenvalues $\tilde D_\nu(\equiv D_\nu/\kappa)$ and atmospheric neutrino
mass-squared difference $\Delta m_{\rm atm}^2$ as follows:
\begin{align}
({\rm NH}):\  \kappa^2= \frac{|\Delta m_{\rm atm}^2|}{\tilde D_{\nu_3}^2-\tilde D_{\nu_1}^2},
\quad
({\rm IH}):\  \kappa^2= \frac{|\Delta m_{\rm atm}^2|}{\tilde D_{\nu_2}^2-\tilde D_{\nu_3}^2}, \label{eq:kappa}
 \end{align}
where 
NH and IH respectively stand for the normal
and inverted hierarchies. Subsequently, the solar neutrino mass-squared difference is determined by the relation
\begin{align}
\Delta m_{\rm sol}^2= {\kappa^2}({\tilde D_{\nu_2}^2-\tilde D_{\nu_1}^2}).
 \end{align}
This value should be within the allowed range of the experimental value. In numerical analysis, we will adopt NuFit 5.2~\cite{Esteban:2020cvm} to experimental ranges of neutrino observables. 
The {effective mass for} neutrinoless double beta decay can be written by 
\begin{align}
\langle m_{ee}\rangle=\kappa|\tilde D_{\nu_1} \cos^2\theta_{12} \cos^2\theta_{13}+\tilde D_{\nu_2} \sin^2\theta_{12} \cos^2\theta_{13}e^{i\alpha_{ 21}}+\tilde D_{\nu_3} \sin^2\theta_{13}e^{i(\alpha_{31}-2\delta_{CP})}|.
\end{align}
%

A predicted value of $\langle m_{ee} \rangle$ is constrained by the current KamLAND-Zen data and could be measured in future~\cite{KamLAND-Zen:2024eml}.
Currently the upper bound is found as $\langle m_{ee}\rangle<(36-156)$ meV at 90 \% confidence level where the range of the bound is due to the use of different method in estimating nuclear matrix elements. 
Furthermore the sum of neutrino masses is constrained by the minimal standard cosmological model
$\Lambda$CDM $+\sum D_{\nu}$ that provides the upper bound $\sum D_{\nu}\le$ 120 meV~\cite{Vagnozzi:2017ovm, Planck:2018vyg}, although it becomes weaker if the data are analyzed in the context of extended cosmological models~\cite{ParticleDataGroup:2014cgo}.
Recently, DESI and CMB data combination provides more stringent upper bound on the sum as $\sum D_{\nu}\le$ 72 meV~\cite{DESI:2024mwx}. 
%
These two observable $\langle m_{ee}\rangle$ and $\sum D_{\nu}$ are also discussed as our prediction in the numerical analysis.
}
Direct search for neutrino mass is done by the Karlsruhe Tritium Neutrino (KATRIN) experiment~\cite{KATRIN:2021uub},
which is the first sub-eV sensitivity on $m_{\nu_e}^2=(0.26\pm0.34)$ eV$^2$ at 90 \% CL.
Here,
$m_{\nu_e}^2$
is defined by
\begin{align}
\kappa^2\left[(\tilde D_{\nu_1} c_{13}c_{12})^2+(\tilde D_{\nu_2} c_{13}s_{12})^2+(\tilde D_{\nu_3} s_{13})^2\right].
\end{align}
We also take this experiment in account in our numerical analysis below.

{

\subsection{Charged Lepton Flavor Violation}

Here we briefly discuss charged lepton flavor violation (CLFV) in the model.
The CLFVs can be induced via Yukawa interactions which are the first two terms in the RHS of Eq.~\eqref{eq:lpy}.
Dominant contributions come from one-loop diagram generating dipole operator~\cite{}; 
\begin{equation}
\mathcal{L}_{\rm eff} = \frac{C_L^{ji}}{M^2} \frac{v}{\sqrt{2}} \overline{\ell_j} \sigma_{\mu \nu} P_L \ell_i F^{\mu \nu} + \frac{C_R^{ji}}{M^2} \frac{v}{\sqrt{2}} \overline{\ell_j} \sigma_{\mu \nu} P_R \ell_i F^{\mu \nu},
\end{equation}
where $F^{\mu \nu}$ is the electromagnetic field strength tensor and $M$ indicate the heavy particle mass inside loop diagrams ($M =M_{L'}$ in our case). 
The $C_{L,R}^{ji}$ is a Wilson coefficient obtained by calculating one-loop diagram. The order of Wilson coefficient is roughly estimated as $C_{L,R} \sim \frac{y^2}{(4 \pi)^2} \frac{m_{\ell_i}}{v} I_{\rm loop} $ where $I_{\rm loop}$ is a $\mathcal{O}(1)$ factor from loop integration and $y$ represents a Yukawa coupling inside a loop diagram.
We then consider $\mu \to e \gamma$ process that would provide the strongest CLFV constraints. From the effective interaction, we can estimate 
\begin{align}
Br(\mu \to e\gamma) \sim & \frac{12 \sqrt{2} \pi}{G_F^2 M^4} \left[ \frac{y^2}{(4 \pi)^2} \frac{m_{\mu}}{v} I_{\rm loop} \right]^2 \nn \\
\sim & 10^{-5} y^4 \left( \frac{\rm TeV}{M} \right)^4,
\end{align}
where $G_F$ is the Fermi constant, and we assumed $I_{\rm loop} \sim 1$.
Thus we can avoid the current constraint~\cite{}, $Br(\mu \to e \gamma) < 1.5 \times 10^{-13}$, by choosing $M \gtrsim 100$ TeV even if Yukawa couplings are $\mathcal{O}(1)$.
In our numerical analysis, therefore, we assume $M_{L'} \gtrsim 100$ TeV and do not discuss CLFVs in detail.
}

{
\section{Mass matrices around the fixed points}
{Here, we discuss modular forms at nearby fixed points, $\tau= \{i,\ \omega\}$, and show concrete structure of mass and mixing matrices in terms of each fixed point and its deviation. }
\subsection{$\tau\simeq i$}
In case of $\tau\simeq i$, we can expand modular Yukawa couplings in terms of $\epsilon\equiv \tau-i$~\cite{Abe:2024tox}.
Here, we expand $Y_3^{(2)}$ up to $\epsilon$ as follows:
\begin{align}
y_1&\approx Y_0 +i \left( \frac{c_{20}}{\sqrt2}+\frac{c'_{20}}{\sqrt6} -\frac{c_{21}}{2\sqrt3}\right) \epsilon +{\cal O}(\epsilon^2) ,\\
y_2&\approx (1-\sqrt3)Y_0 +i \left( \frac{c_{20}}{\sqrt2}+\frac{c'_{20}}{\sqrt6} -\frac{c_{21}}{2\sqrt3}\right) \epsilon +{\cal O}(\epsilon^2) ,\\
y_3&\approx (-2+\sqrt3)Y_0 - i \left( \sqrt{\frac23}c'_{20} +\frac{c_{21}}{2\sqrt3} \right) \epsilon +{\cal O}(\epsilon^2) ,
\end{align}
$c_{20}\simeq 1.252$, $c'_{20}\approx0.336$, and $c_{21}\approx3.049$.
Since $m'$ as well as $m_e$ is represented via $Y_3^{(2)}$, the form of charged-lepton mass matrix proportional to
\begin{align}
 \left(\begin{array}{ccc} 
y_1 & y_3 & y_2 \\
y_2 & y_1 & y_3\\
y_3& y_2 & y_1  \end{array} \right) \simeq 
Y_0 
\left(\begin{array}{ccc} 
1 & -2+\sqrt3 & 1-\sqrt3 \\
1-\sqrt3 & 1 & -2+\sqrt3\\
-2+\sqrt3 & 1-\sqrt3 & 1  \end{array} \right)
+%
\frac{i}{2\sqrt3} \left(\begin{array}{ccc} 
a_0 & c_0 & b_0 \\
b_0 &a_0 & c_0\\
c_0 & b_0 &a_0  \end{array} \right) \epsilon
+{\cal O}(\epsilon^2),
\end{align}
where $a_0=\sqrt6 c_{20}+\sqrt2 c'_{20}-c_{21}$, $b_0=-\sqrt6 c_{20}+\sqrt2 c'_{20}-c_{21}$,
and $c_0=-2\sqrt2 c'_{20}-c_{21}$.
Thus, the $V_{e_L}$ to diagonalize the dominant matrix, which is denoted by $V_{e_L}^0$, is found as
\begin{align}
V_{e_L}^0=
 \left(
 \begin{array}{ccc} 
\frac1{\sqrt3} & \frac1{\sqrt2} & \frac1{\sqrt6} \\
\frac1{\sqrt3} & -\frac1{\sqrt2} & \frac1{\sqrt6}\\
\frac1{\sqrt3}& 0 & -\sqrt{\frac23}  \end{array} 
\right)
 \left(
 \begin{array}{ccc} 
1 &0 & 0 \\
0 & \cos\theta & -\sin\theta\\
0 & \sin\theta & \cos\theta  \end{array} 
\right) .
\end{align}
$\sin\theta$ is determined by solving the following condition:
\begin{align}
(-2+\sqrt3) (2 a_e^2 - b_e^2 -c_e^2) \cos2\theta+(-3+2\sqrt3) (b_e^2 -c_e^2) \sin2\theta =0.
\end{align}
At the zeroth order of charged-lepton mass eigenvalues; $m_e^0,m^0_\mu,m^0_\tau$, 
we have $m_e^0=0$ where we abbreviate the other two mass eigenvalues due to too complicated forms.
It suggests that the electron mass eigenvalue has to rely on the correction terms $\epsilon$.
Hereafter, we redefine $\epsilon_a=\frac{a_e}{M_1}$, $\epsilon_b=\frac{b_e}{M_2}$, and $\epsilon_c=\frac{c_e}{M_3}$.

Then, the neutrino mass matrix at the leading order is given by
\begin{align}
m_\nu^0 \simeq
Y_0^2 \frac{v_\Delta y_\Delta}{\sqrt2}
\left[
 \tilde y'_\Delta \epsilon_b \epsilon_c \left(
 \begin{array}{ccc} 
-10+6\sqrt3 & 2-\sqrt3 & 8-5\sqrt3 \\
2-\sqrt3 & 2-2\sqrt3 & -4+3\sqrt3\\
8-5\sqrt3 & -4+3\sqrt3 & 2(-2+\sqrt3)  \end{array} 
\right) 
+
\left(
 \begin{array}{ccc} 
1& -2+\sqrt3 & 1-\sqrt3 \\
-2+\sqrt3 & 7-4\sqrt3 & -5+3\sqrt3\\
1-\sqrt3 & -5+3\sqrt3 & 4-2\sqrt3  \end{array} 
\right) \epsilon_a^2
\right]
.
\end{align}
The mixing matrix is given by
\begin{align}
U_{\nu}^0=  
\left(
 \begin{array}{ccc} 
\frac1{\sqrt3} & \frac1{\sqrt2} & \frac1{\sqrt6} \\
\frac1{\sqrt3} & -\frac1{\sqrt2} & \frac1{\sqrt6}\\
\frac1{\sqrt3}& 0 & -\sqrt{\frac23}  \end{array} 
\right)
 \left(
 \begin{array}{ccc} 
1 &0 & 0 \\
0 & \frac1{\sqrt2} & - \frac1{\sqrt2} \\
0 & \frac1{\sqrt2}  &  \frac1{\sqrt2}   \end{array} 
\right) .
\end{align}
Then, the mass eigenvalues at leading order are found as
\begin{align}
D_{\nu_1}^0 =0
,\
D_{\nu_2}^0=9\sqrt{7-4\sqrt3}Y_0^2  \tilde y'_\Delta \epsilon_b \epsilon_c,
,\
D_{\nu_3}^0=3\sqrt{7-4\sqrt3}Y_0^2 (2\epsilon^2+ \tilde y'_\Delta \epsilon_b \epsilon_c).
\end{align}
The PMNS at the leading order;  $U^0$, is given by
\begin{align}
U^0=
\left(
 \begin{array}{ccc} 
1 & 0 & 0 \\
0 & \frac{\cos\theta+\sin\theta}{\sqrt2} &\frac{-\cos\theta+\sin\theta}{\sqrt2}\\
0 & \frac{\cos\theta-\sin\theta}{\sqrt2} &\frac{\cos\theta+\sin\theta}{\sqrt2}  \end{array} 
\right).
\end{align}
As can be seen from the above form, 1-2 mixing and 1-3 mixing angles have to be generated via correction terms.
However 1-2 mixing is not so small via experiments and we could suppose that $\tau=i$ would be disfavored.
In fact we have checked that there is no solutions at nearby $\tau=i$ unless $\Delta \chi^2$ is greater than $10^3$.
Thus, we concentrate on the region at nearby $\tau=\omega$.

\subsection{$\tau\simeq \omega$}
In case of $\tau\simeq \omega$, we can expand modular Yukawa couplings in terms of $\epsilon\equiv \tau-\omega$~\cite{Abe:2024tox}.
Here, we expand $Y_3^{(2)}$ up to $\epsilon$ follows:
\begin{align}
y_1&\approx \frac23 \{ b_0 -\frac{i}{\sqrt3} (b_1-2b_0) \epsilon +{\cal O}(\epsilon^2) \},\\
y_2&\approx \frac{\omega}3 \{ 2b_0 +\frac{i}{\sqrt3} (b_1+4b_0) \epsilon +{\cal O}(\epsilon^2) \},\\
y_3&\approx - \frac{\omega^2}3 \{b_0 +\frac{2i}{\sqrt3} (b_1+b_0) \epsilon +{\cal O}(\epsilon^2) \},
\end{align}
$b_0\simeq 1.423$ and $b_1\approx3.673$.
Since $m'$ as well as $m_e$ is represented via $Y_3^{(2)}$, the form of charged-lepton mass matrix proportional to
\begin{align}
 \left(\begin{array}{ccc} 
y_1 & y_3 & y_2 \\
y_2 & y_1 & y_3\\
y_3& y_2 & y_1  \end{array} \right) \simeq 
\frac{b_0}3 \left(\begin{array}{ccc} 
2 & -\omega^2 & 2\omega \\
2\omega & 2 & -\omega^2\\
-\omega^2 & 2\omega & 2  \end{array} \right)
+%
\frac{i}{3\sqrt3} \left(\begin{array}{ccc} 
2(2 b_0-b_1) & -2(b_0+b_1)\omega^2 & (4b_0+b_1)\omega \\
(4b_0+b_1)\omega & 2(2 b_0-b_1) & -2(b_0+b_1)\omega^2\\
-2(b_0+b_1)\omega^2 & (4b_0+b_1)\omega & 2(2 b_0-b_1)  \end{array} \right) \epsilon
+{\cal O}(\epsilon^2).
\end{align}
Thus, the $V_{e_L}$ to diagonalize the dominant matrix, which is denoted by $V_{e_L}^0$, is found as
\begin{align}
V_{e_L}^0=
\frac13 \left(
 \begin{array}{ccc} 
2\omega & -2\omega & -\omega \\
-\omega^2 & -2\omega^2 & 2\omega^2\\
2& 1 & 2  \end{array} 
\right) .
\end{align}
And the zeroth order of charged-lepton mass eigenvalues; $m_e^0,m^0_\mu,m^0_\tau$, are given by
\begin{align}
& m_e^0 = \sqrt{\frac{vv'}2}\frac{a_e f_1}{M_1} b_0,\
 m_\mu^0 = \sqrt{\frac{vv'}2}\frac{b_e f_2}{M_2} b_0,\
 m_\tau^0 = \sqrt{\frac{vv'}2}\frac{c_e f_2}{M_3} b_0.
\end{align}
It implies that we have the following hierarchy from the experimental masses for the charged-leptons 
\begin{align}
\frac{a_e}{M_1}\ll \frac{b_e}{M_2} < \frac{c_e}{M_3}.
\end{align}
Hereafter, we redefine $\epsilon_a=\frac{a_e}{M_1}$, $\epsilon_b=\frac{b_e}{M_2}$, and $\epsilon_c=\frac{c_e}{M_3}$.

Then, the neutrino mass matrix is expanded via $\epsilon_a$ in the limit of $\epsilon=0$, and its form is given by
\begin{align}
m_\nu^0 \simeq
b_0^2 \frac{v_\Delta y_\Delta}{9\sqrt2}
\left[
 \tilde y'_\Delta \epsilon_b \epsilon_c \left(
 \begin{array}{ccc} 
-4& 2\omega^2 & 5\omega \\
2\omega^2 & 8\omega & 2\\
5\omega & 2 & -4 \omega^2  \end{array} 
\right) 
+
\left(
 \begin{array}{ccc} 
4& -2\omega^2 & 4\omega \\
-2\omega^2 & \omega & -2\\
4\omega & -2 & 4 \omega^2  \end{array} 
\right) \epsilon_a^2
\right]
.
\end{align}
The mixing matrix is given by
\begin{align}
U_{\nu}^0= \frac13
\left(
 \begin{array}{ccc} 
2 \omega & -\omega^2  & 2  \\
-\frac{3\omega}{\sqrt2}   & 0 & \frac{3}{\sqrt2} \\
\frac{\omega^2}{\sqrt2}   & 2\sqrt2 &\frac{\omega}{\sqrt2}   
\end{array} 
\right) .
\end{align}
Then, the mass eigenvalues are found as
\begin{align}
D_{\nu_1}^0 = b_0^2 \epsilon_a^2
,\
D_{\nu_2}^0=b_0^2  \tilde y'_\Delta \epsilon_b \epsilon_c,
,\
D_{\nu_3}^0=  b_0^2  \tilde y'_\Delta \epsilon_b \epsilon_c.
\end{align}
As a result, we have two large mixing angles of PMNS at the leading order.
Below, we will numerically show there exist solutions at nearby $\tau=\omega$.

}

\section{Numerical analysis and phenomenology}
\label{sec:III}
In this section, we perform chi square numerical analysis to fit the neutrino data and discuss several features of our model
{where we refer to NuFit 5.2~\cite{Esteban:2020cvm}.
The $\Delta \chi^2$ value is calculated as 
\begin{equation}
\Delta \chi^2 = \sum_{i}  \left( \frac{O_i^{\rm obs} - O_i^{\rm th}}{\delta O_i^{\rm exp}} \right)^2, \label{eq:chi-square}
\end{equation}
where $O_i^{\rm obs (th)}$ is observed (theoretically obtained) value of corresponding observables and $\delta O_i^{\rm exp}$ expresses experimental error.
Through this analysis, Dirac CP phase and Majorana phase are considered as output parameters since any values are allowed by experiments at 3$\sigma$ confidence level (C.L.).
Therefore, we make the use of five reliable observables [$\Delta m_{\rm atm}^2,\ \Delta m_{\rm sol}^2,\ \sin\theta_{12},\ \sin\theta_{23},\ \sin\theta_{13}$].
\footnote{{Since $\sin\theta_{23}$ is far from the Gaussianity, we directly take the data from Nufit 5.2.}}
Five degrees of freedom lead us to  
$\Delta \chi^2=5.88$ at 1$\sigma$, $\Delta \chi^2=11.3139$ at 2$\sigma$, $\Delta \chi^2=18.2053$ at 3$\sigma$, $\Delta \chi^2=26.7663$ at 4$\sigma$, and $\Delta \chi^2=37.0948$ at 5$\sigma$.
}
Furthermore,  we concentrate on the region at nearby two representative fixed points $\tau=\{i,\omega\}$ and search for the allowed region within the range of $5\sigma$ with five degrees of freedom. However, we could not find any solutions to satisfy the neutrino oscillation data in case of $\tau=i$. Thus, we focus on discussing features in case of $\tau=\omega$ below.

{ In addition to modulus $\tau \sim \omega$, we have 13 relevant free parameters for lepton mass matrices that are
$\{v', v_\Delta,  M_1, M_2, M_3, a_e, a_b, a_c, f_1, f_2, f_3, y_\Delta, y'_\Delta\}$ where only $y'_\Delta$ is complex. 
Three parameters, $\{a_e, b_e, c_e\}$, are fixed to fit charged lepton masses via Eq.~\eqref{eq:l-cond}. Six parameters, $\{y_\Delta, v_\Delta, v', M_1, f_1, y_\Delta \}$, are factored out and they are only related to overall scale of mass matrices.
Then neutrino mass squared differences and mixings are fitted by remaining rescaled free parameters that are $\{\tilde{f}_2, \tilde{f}_3, \tilde{M}_2, \tilde{M}_3, \tilde{y}_\Delta \}$ where $\tilde{f}_{2,3} \equiv f_{2,3}/f_1$ and $\tilde{M}_{2,3} \equiv M_{2,3}/M_1$.
}
We randomly select all the rescaled input parameters within the ranges of $[10^{-3}-10^{3}]$ except for $|\tilde y'_\Delta|$ that is  $[10^{-5}-1]$ through our numerical analysis.
{ We take mass scale as $m' > 1$ TeV, $M_{1,2,3} > 100$ TeV to satisfy our assumption regarding hierarchy of mass parameters and CLFV constraints discussed in the previous section. In addition, we implicitly satisfy $m'/M_{1}\lesssim 10^{-3}$ to evade  non-unitarity bounds.}

\begin{figure}[tb]
\begin{center}
\includegraphics[width=50.0mm]{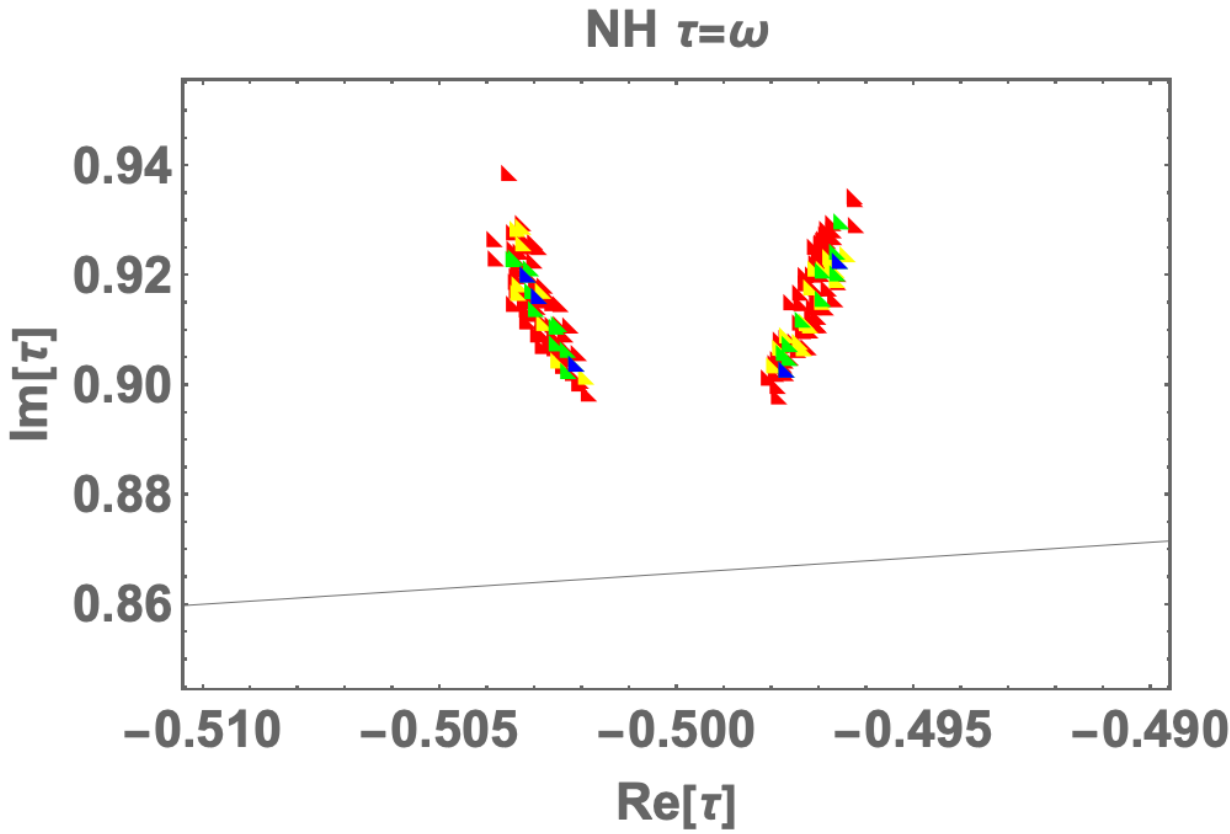} \quad
\includegraphics[width=50.0mm]{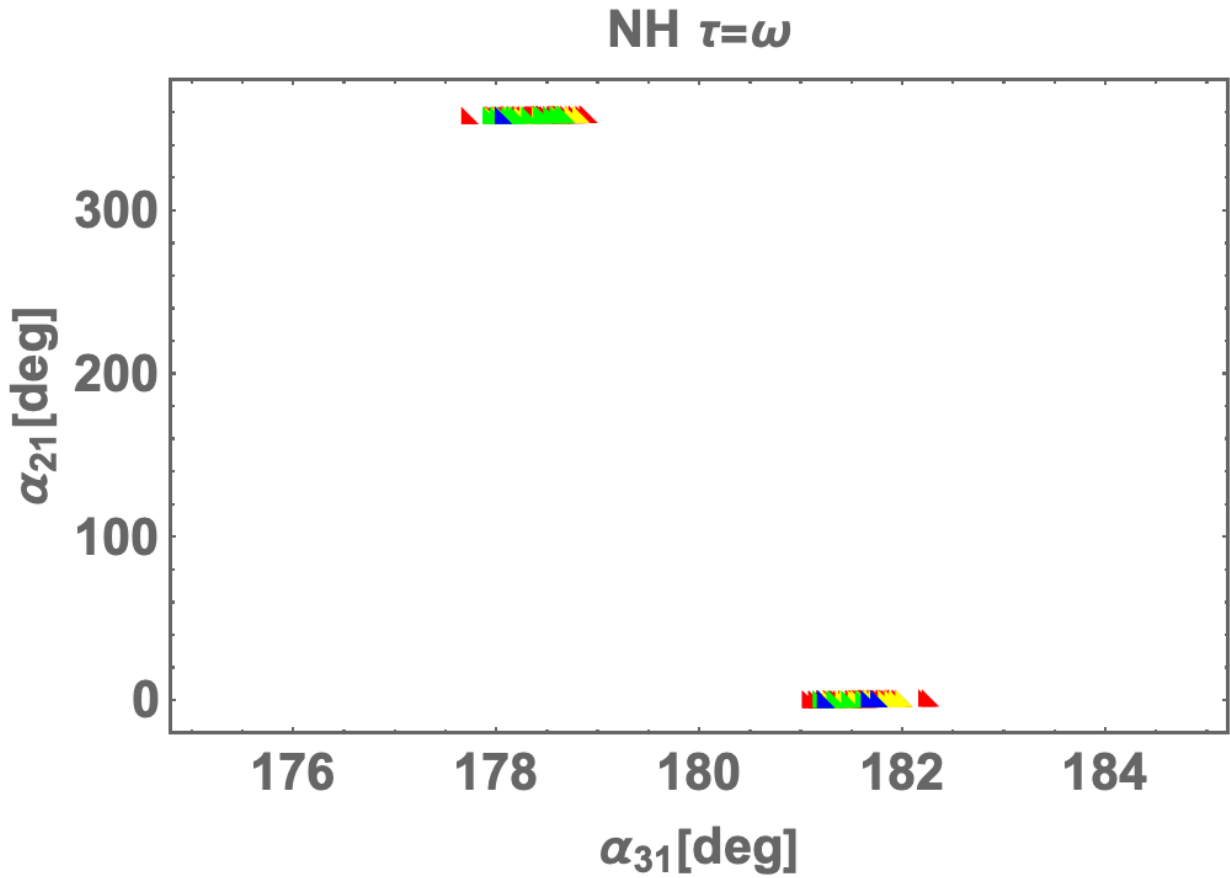} \quad
\includegraphics[width=50.0mm]{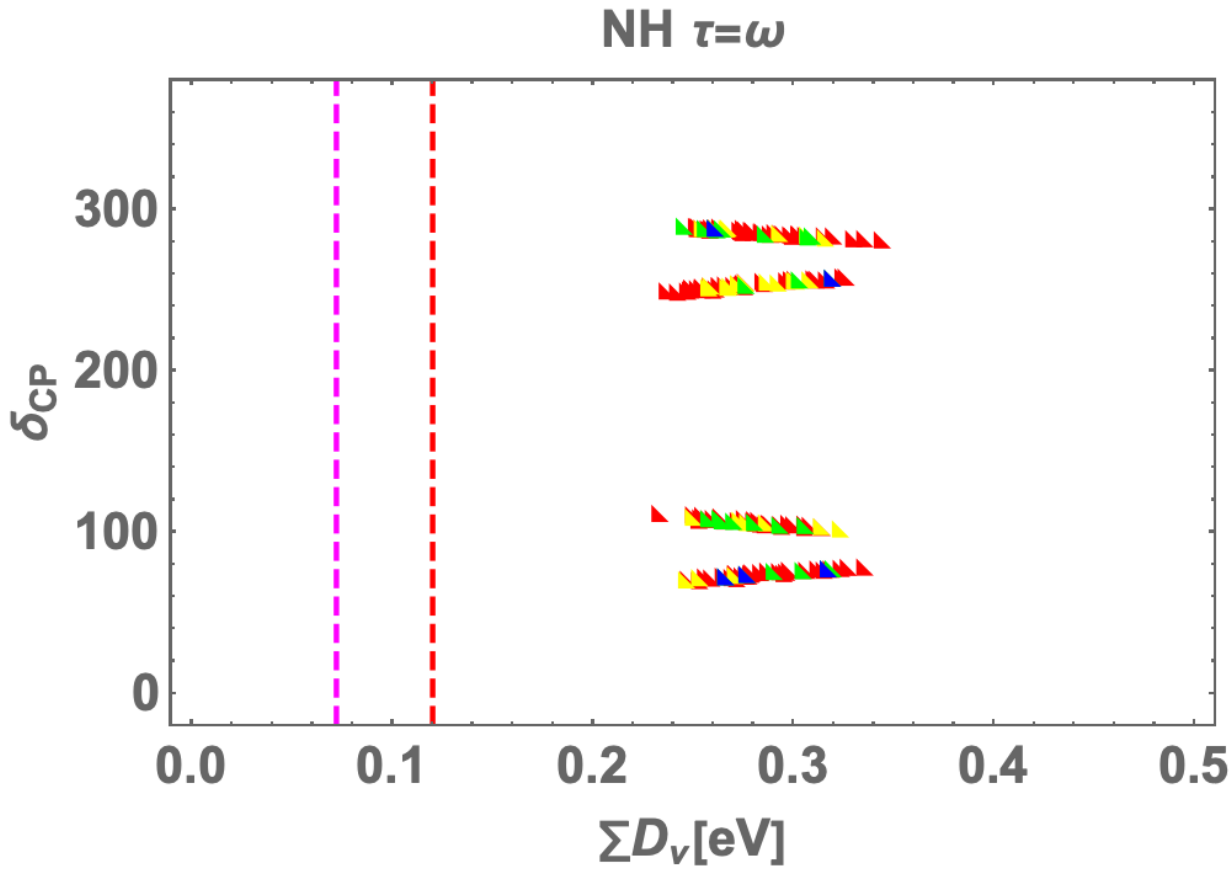} 
\caption{Numerical $\Delta\chi^2$ analyses in case of NH at nearby $\tau=\omega$, where the blue color represents the range of $0-1$, the green $1-2$, the yellow $2-3$, and the red $3-5$ {$\sigma$ standard deviation estimated by} $\sqrt{\Delta\chi^2}$. 
The left figure shows the allowed range of $\tau$ where the black solid line is the boundary of the fundamental domain at $|\tau|=1$. 
The central figure represents the allowed range of Majorana phases. The right figure demonstrates
the allowed Dirac CP phase $\delta_{\rm  CP}$ in terms of sum of neutrino masses $\sum D_\nu$, where the red dotted line is the cosmological upper bound while the pink dotted line is upper bound on the DESI and CMB data.  }
  \label{fig:omega_nh1}
\end{center}\end{figure}
%
\subsection{NH}
In this subsection, we discuss the allowed regions in case of NH.
%
In Figs.~\ref{fig:omega_nh1}, the blue color represents the range of $0-1$, the green $1-2$, the yellow $2-3$, and the red $3-5$ {$\sigma$ standard deviation estimated by} { $\Delta\chi^2$}. 
The left figure shows the allowed range of $\tau$ where the black solid line is the boundary of the fundamental domain at $|\tau|=1$. 
It suggests that we have enough solutions at nearby $\tau=\omega$.
The central figure represents the allowed range of Majorana phases $\alpha_{21}$ and $\alpha_{31}$.
{We find}  $\alpha_{21}$ is almost zero degree but $\alpha_{31}$ is allowed by the ranges [$177.5^\circ - 178.5^\circ$] and [$181^\circ - 182.5^\circ$].

The right figure of Fig.~\ref{fig:omega_nh1} demonstrates
the allowed region of Dirac CP phase $\delta_{\rm  CP}$ in terms of sum of neutrino masses $\sum D_\nu$, where the red dotted line is the cosmological upper bound while the pink dotted line is upper bound from {the combination of} the DESI and CMB data. 
 $\delta_{\rm  CP}$ is allowed by the ranges of  [$80^\circ - 100^\circ$] and [$240^\circ - 300^\circ$].
 $\sum D_\nu$ is allowed by the range of [$0.22\ {\rm eV}- 0.35\ {\rm eV}$].

\begin{figure}[tb]
\begin{center}
\includegraphics[width=50.0mm]{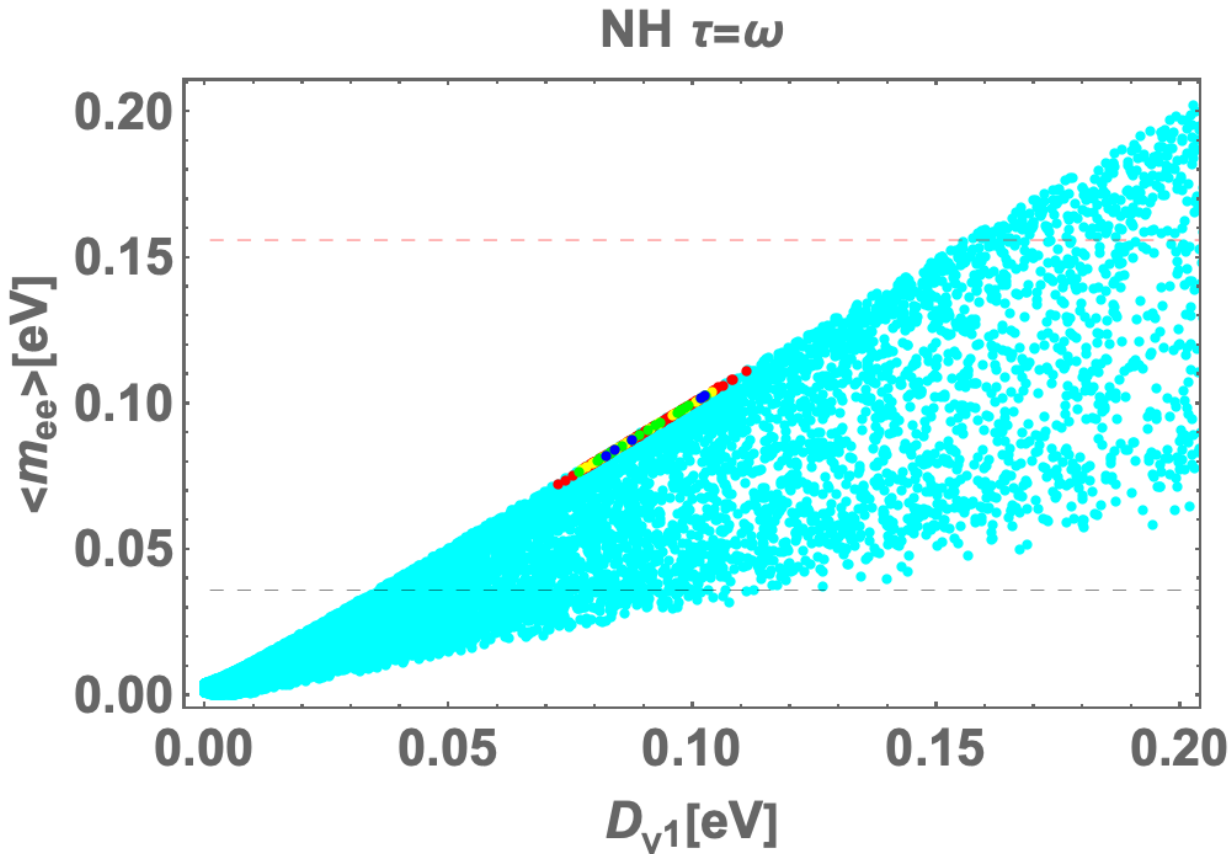} \quad
\includegraphics[width=50.0mm]{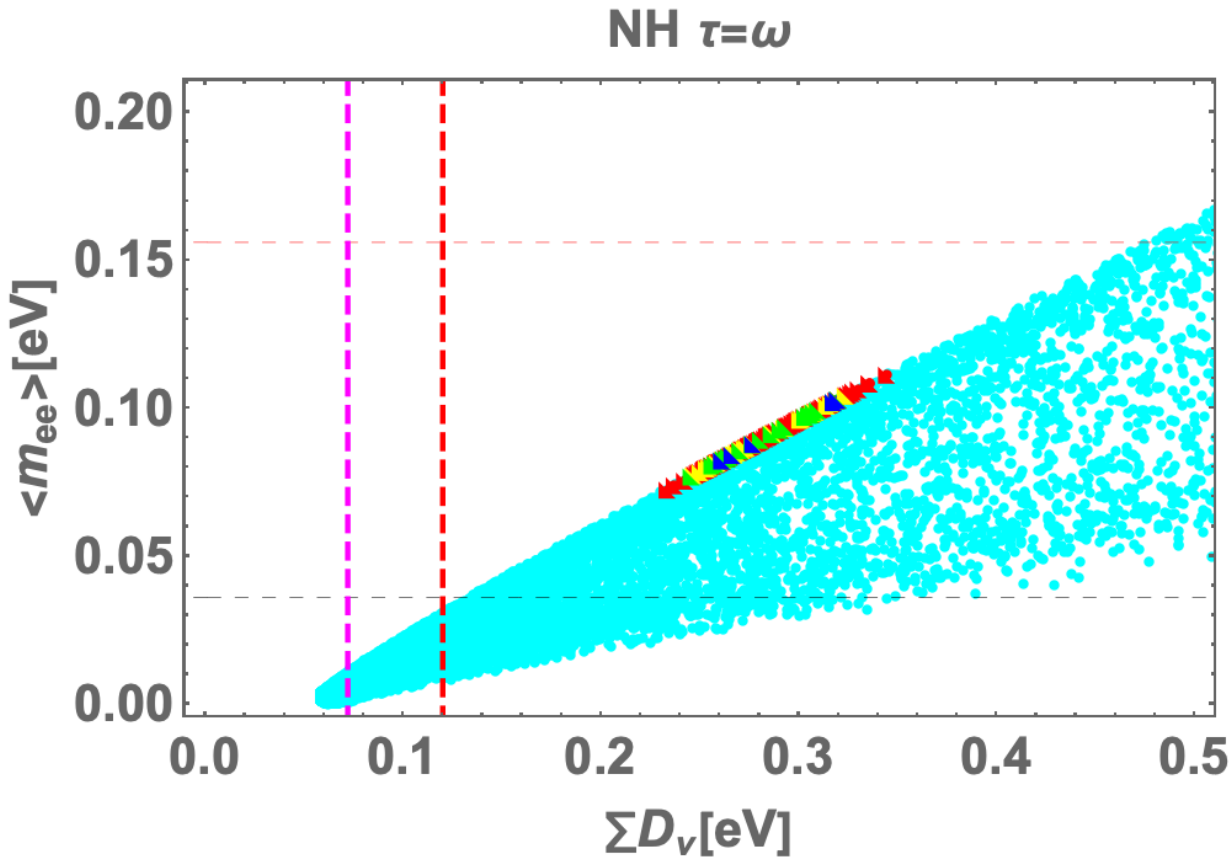} \quad
\includegraphics[width=50.0mm]{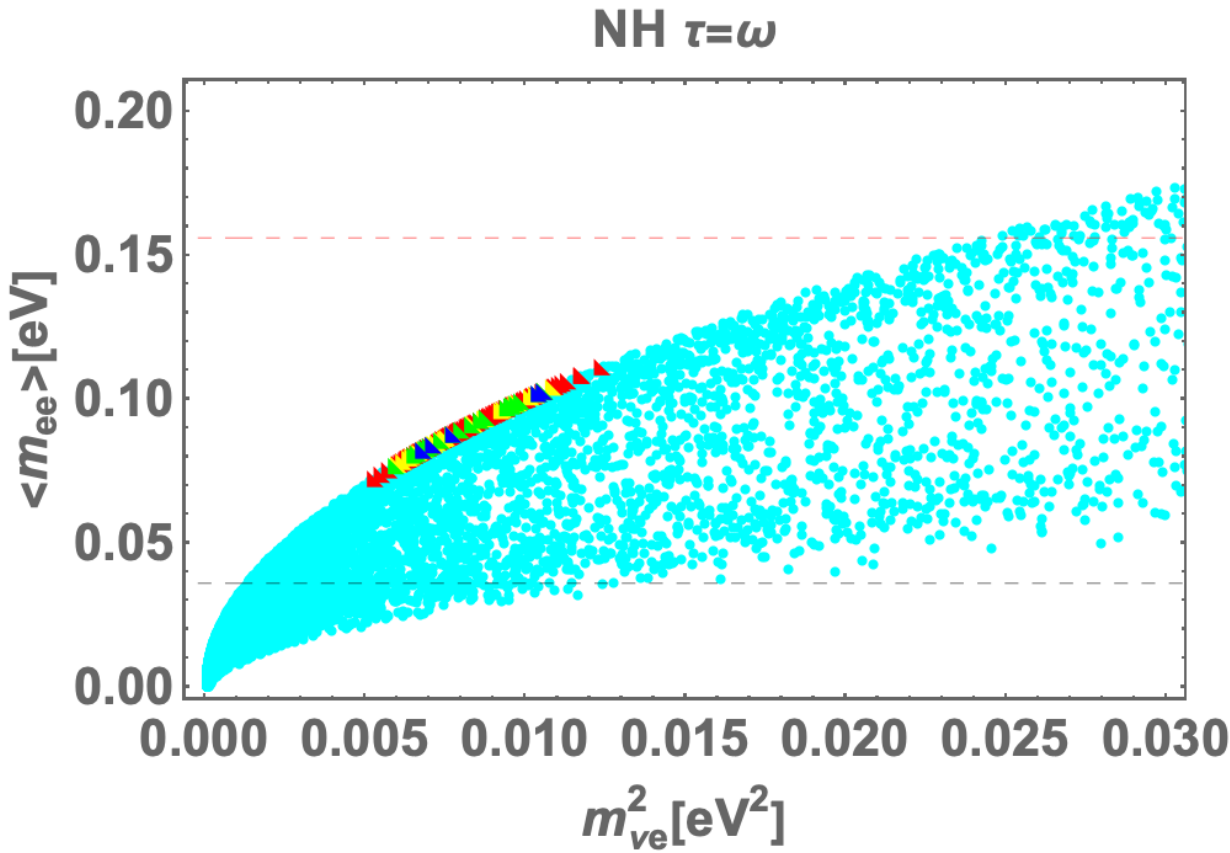} 
\caption{Numerical $\Delta\chi^2$ analyses in case of NH at nearby $\tau=\omega$, where color legends are the same as  the ones of Fig.~\ref{fig:omega_nh1}.
These figures show the allowed range of the neutrinoless double beta decay $\langle m_{ee}\rangle$ in terms of the lightest neutrino mass $D_{\nu_1}$ (left), $\sum D_\nu$, and $m_{\nu_e}^2$ (right) in unit of eV. {Cyan region is allowed by experimental result of Nufit 6.0.}}
  \label{fig:omega_nh2}
\end{center}\end{figure}
%
In Fig.~\ref{fig:omega_nh2}, we show some predictions in the case of NH at nearby $\tau=\omega$, where color legends are the same as  the ones of Fig.~\ref{fig:omega_nh1}.
The red and gray dotted lines respectively correspond to upper and lower bounds of $\langle m_{ee}\rangle$ experiment at 90 \% {confidence} level. {Cyan region is allowed by experimental result of Nufit 6.0.}
These figures show the allowed range of the neutrinoless double beta decay $\langle m_{ee}\rangle$ in terms of the lightest neutrino mass $D_{\nu_1}$ (left), $\sum D_\nu$, and $m_{\nu_e}^2$ (right) in unit of eV.
These suggest each of allowed region as follows:
$\langle m_{ee}\rangle=[0.07\ {\rm eV}-0.12\ {\rm eV}]$ that is completely within the range of the experiment.
The lightest neutrino mass spans over the range of $D_{\nu_1}=[0.07\ {\rm eV}-0.11\ {\rm eV}]$ whose range is similar to the one of neutrinoless double beta decay.
Even though our $\sum D_\nu$ does not satisfy the cosmological bound $0.12$ [eV], $m_{\nu_e}^2$ is totally safe to the experimental upper bound.

%
\begin{table}[tb]
    \setlength\tabcolsep{0.2cm}
    \begin{tabular}{c|c||c|c||c|c}
\hline
        parameter    &  BF & parameter & BF & parameter & BF \\ \hline \hline
          $\tau$ &  $-0.503 + 0.916 i $ & $\tilde y_\Delta$ & $-0.000582005 + 0.00028466 i$  &$\tilde f_2$ & $51.7919$ \\ \hline
       $\tilde f_3$  &8.42753 &  $\tilde M_2$ & $42.6753$ &
 $\tilde M_3$ &   $0.00620328$ \\ \hline
 $\tilde a_e/{\rm GeV}$ &   $0.000361577$ &
$\tilde b_e/{\rm GeV}$ & $0.0761912$ & $\tilde c_e/{\rm GeV}$ & $1.29171$ \\ \hline
$s_{12}$ & 0.547201 &$s_{23}$ & 0.682414 &$s_{13}$ & 0.150202 \\ \hline
$\Delta m^2_{\rm sol}$ & $7.45632\times 10^{-5}~{\rm eV}^2$ &  $\Delta m^2_{\rm atm}$ & $2.52784\times 10^{-3}~{\rm eV}^2$ &
$\delta_{\rm CP}$ & $73.7719^{\circ}$ \\ \hline
$\alpha_{21}$ & $1.19252^{\circ}$ & $\alpha_{31}$ & $181.673^{\circ}$ & $\kappa$ & $48.7~{\rm meV}$ \\ \hline
$m^2_{\nu e}$ & $0.00776348~{\rm eV}^2$ &$\sum D_{\nu}$ & $276.805~{\rm meV}$ & $\langle m_{ee}\rangle$ 
& $87.7089~ {\rm meV}$ \\
\hline
    \end{tabular}
    \caption{\label{tab:BF}%
      Best-fit (BF) parameter values in the NH case corresponding to $\Delta\chi_{\rm min} =1.56$, where $\tilde a_e\equiv \frac{vv'}2 \frac{a_e f_1}{M_1}$, $\tilde b_e\equiv \frac{vv'}2 \frac{b_e f_2}{M_2}$, and $\tilde c_e\equiv \frac{vv'}2 \frac{c_e f_3}{M_3}$.}
\end{table}

\subsection{IH}
%
\begin{figure}[tb]
\begin{center}
\includegraphics[width=50.0mm]{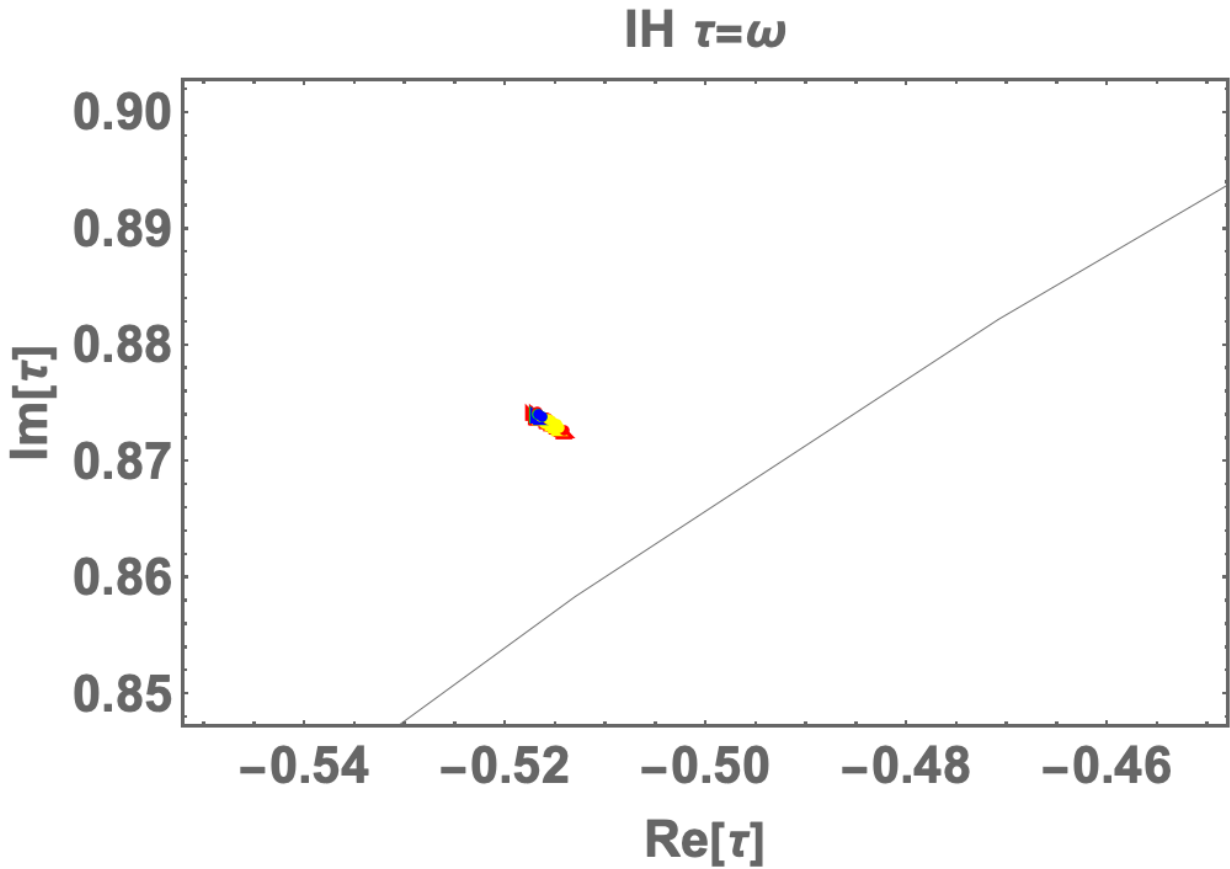} \quad
\includegraphics[width=50.0mm]{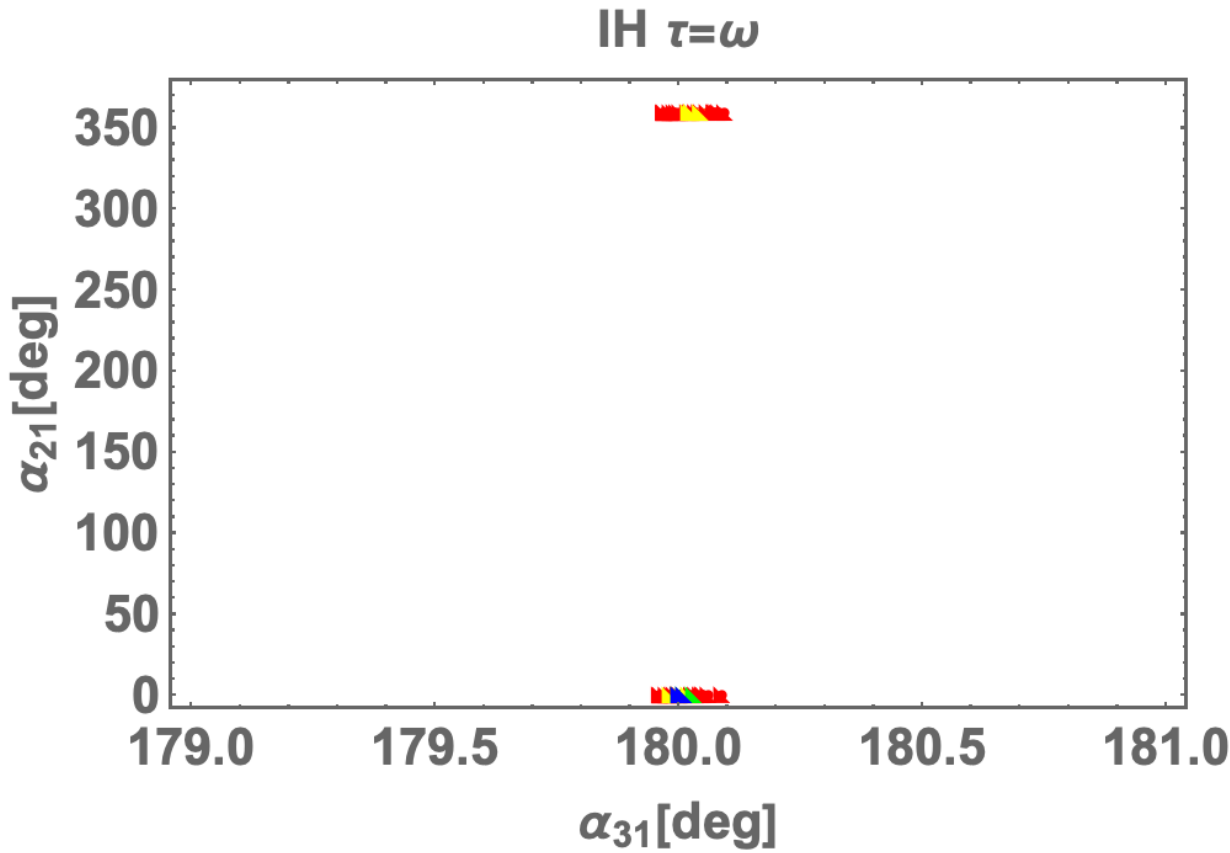} \quad
\includegraphics[width=50.0mm]{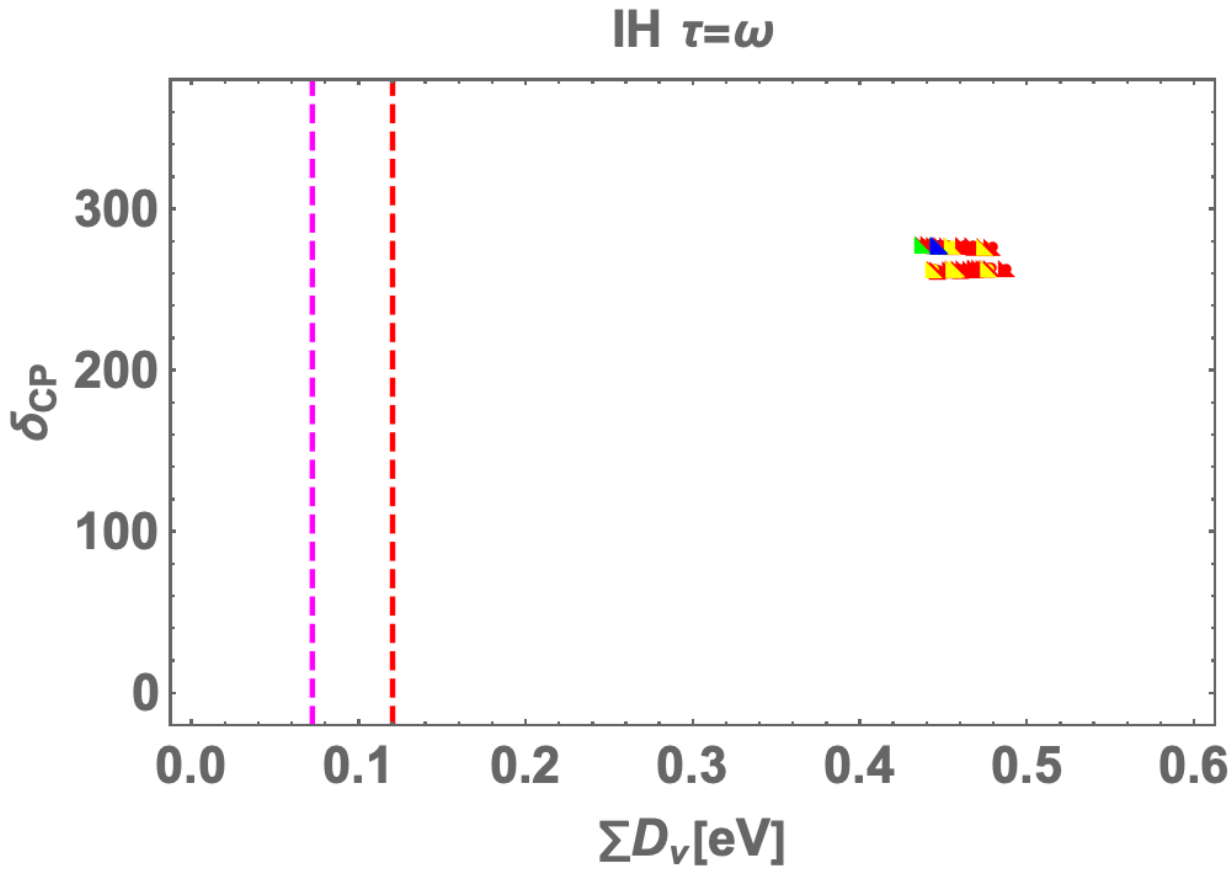} 
\caption{Numerical $\Delta\chi^2$ analyses in case of IH at nearby $\tau=\omega$, where color legends and the captions are the same as  the ones of Fig.~\ref{fig:omega_nh1}.}
  \label{fig:omega_ih1}
\end{center}\end{figure}
%
In this subsection, we discuss the allowed regions in case of IH.
In Figs.~\ref{fig:omega_ih1}, we show the same kinds of figures as in Figs.~\ref{fig:omega_nh1}, and color legends and the captions are the same as  the ones of Fig.~\ref{fig:omega_nh1}.
The left figure suggests that we have enough solutions at nearby $\tau=\omega$, but points are more localized than the case of NH.
The central figure implies that 
 $\alpha_{21}$ is almost zero degree that is the same as the case of NH.
We find $\alpha_{31}$ is localized at nearby 180$^\circ$ that is also similar tendency as the case of NH.

The right figure of Fig.~\ref{fig:omega_ih1} demonstrates
that
 $\delta_{\rm  CP}$ is allowed by the ranges of [$260^\circ - 280^\circ$] while
 $\sum D_\nu$ is allowed by the range of [$0.44\ {\rm eV}- 0.50\ {\rm eV}$] that is a typical tendency.

\begin{figure}[tb]
\begin{center}
\includegraphics[width=50.0mm]{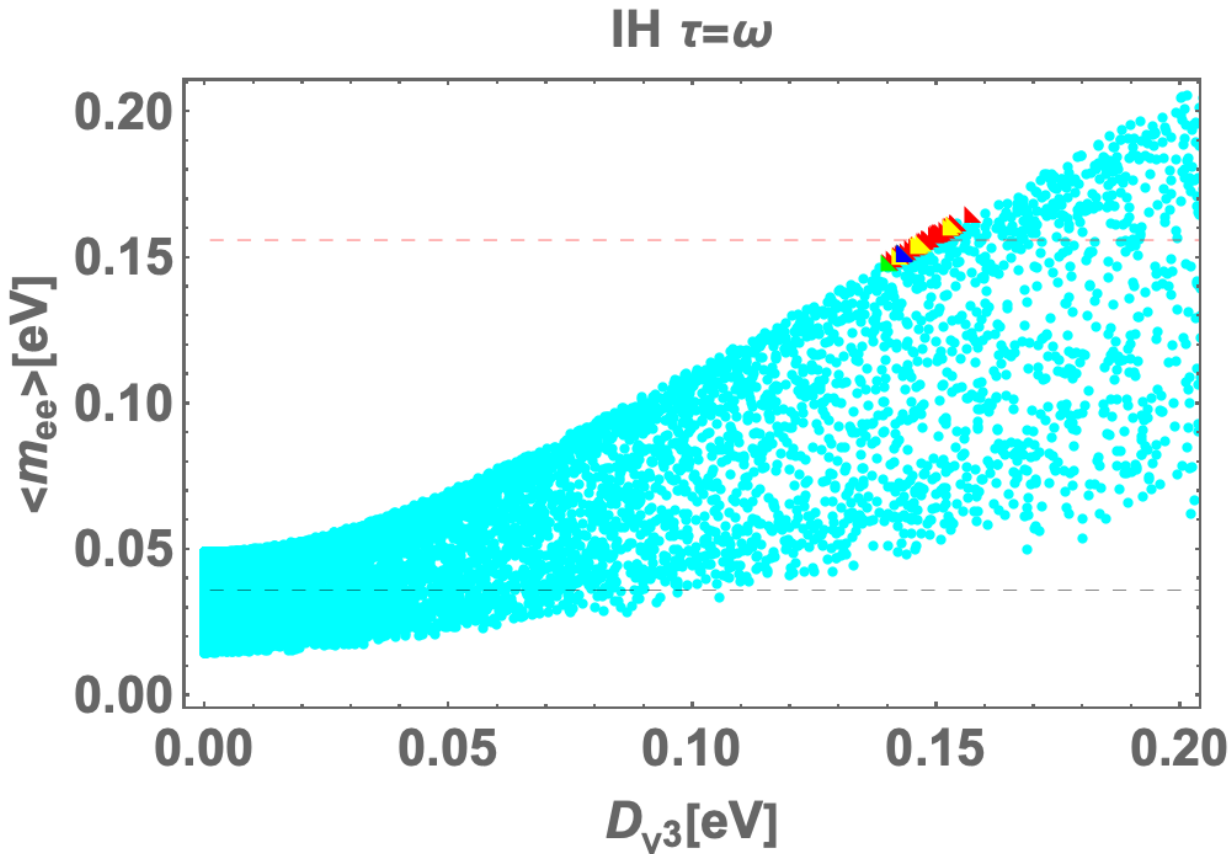} \quad
\includegraphics[width=50.0mm]{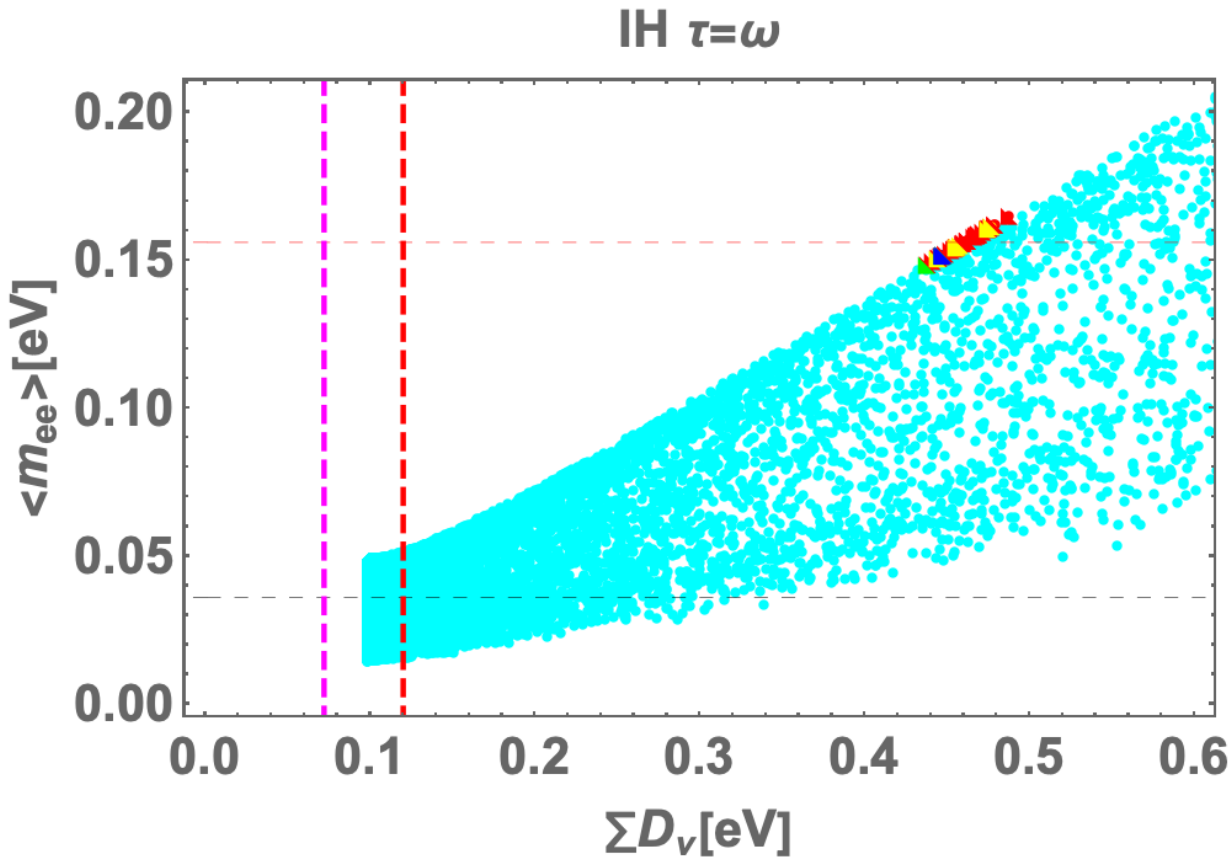} \quad
\includegraphics[width=50.0mm]{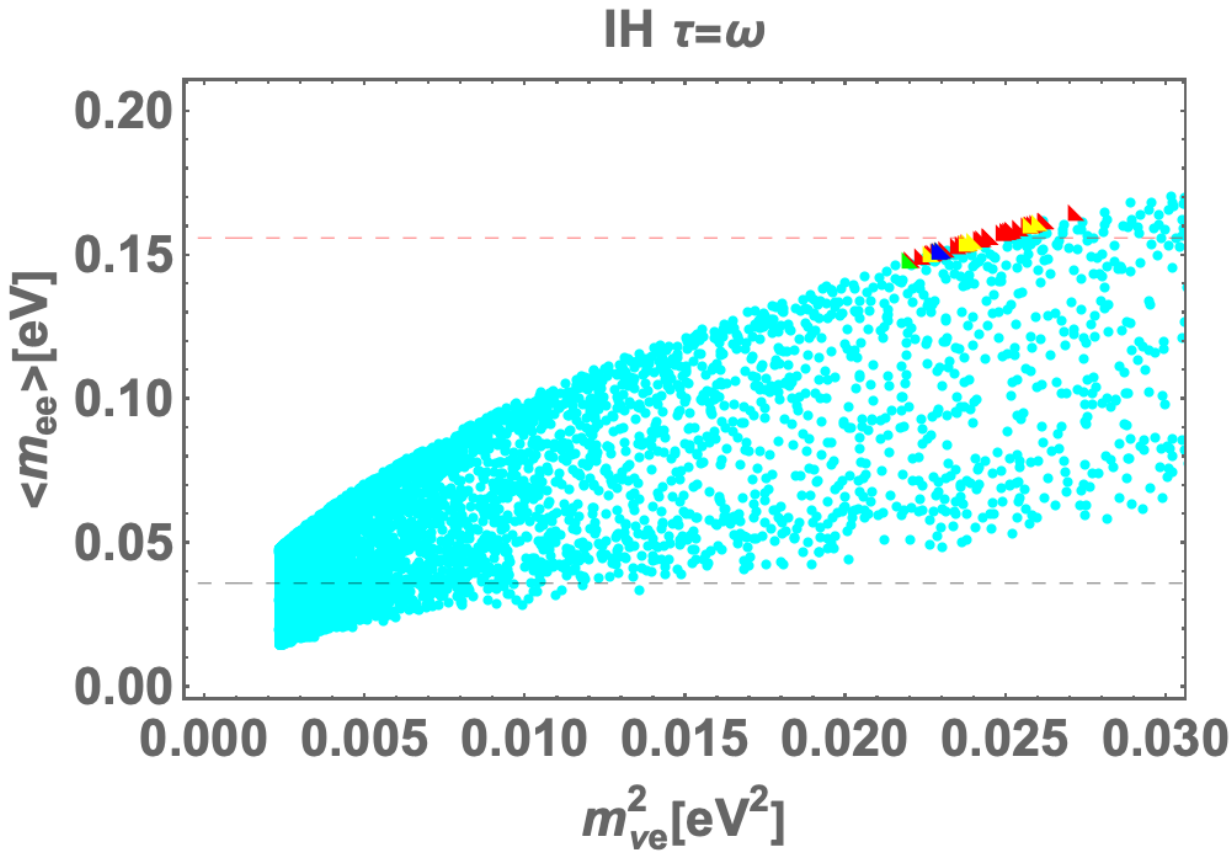} 
\caption{Numerical $\Delta\chi^2$ analyses in case of IH at nearby $\tau=\omega$, where color legends are the same as  the ones of Fig.~\ref{fig:omega_nh1}.
These figures show the allowed range of the neutrinoless double beta decay $\langle m_{ee}\rangle$ in terms of the lightest neutrino mass $D_{\nu_1}$ (left), $\sum D_\nu$, and $m_{\nu_e}^2$ (right) in unit of eV. {Cyan region is allowed by experimental result of Nufit 6.0.}}
  \label{fig:omega_ih2}
\end{center}\end{figure}
%
In Fig.~\ref{fig:omega_ih2}, we show the same kinds of figures as in Fig.~\ref{fig:omega_nh2}, and color legends and the captions are the same as  the ones of Fig.~\ref{fig:omega_nh2}.
%
These suggest each of allowed region as follows:
$\langle m_{ee}\rangle=[0.14\ {\rm eV}-0.16\ {\rm eV}]$ that is nearby the range of upper bound of this experiment. Therefore, this case has high testability {that could be confirmed/excluded in near future}.
The lightest neutrino mass spans over the range of $D_{\nu_3}=[0.14\ {\rm eV}-0.16\ {\rm eV}]$ whose range is similar to the one of neutrinoless double beta decay. The tendency is likely to the case of NH.
$m_{\nu_e}^2$ has the same tendency as the case of NH.
Therefore, it is totally safe to the experimental upper bound.

\begin{table}[tb]
    \setlength\tabcolsep{0.2cm}
    \begin{tabular}{c|c||c|c||c|c}
\hline
        parameter    &  BF & parameter & BF & parameter & BF \\ \hline \hline
          $\tau$ &  $-0.516664 + 0.874007 i $ & $\tilde y_\Delta$ & $-0.0492088 + 0.0215522 i$  &$\tilde f_2$ & $0.213211$ \\ \hline
       $\tilde f_3$  &8.24262 &  $\tilde M_2$ & $15.2601$ &
 $\tilde M_3$ &   $0.00635192$ \\ \hline
 $\tilde a_e/{\rm GeV}$ &   $0.00034419$ &
$\tilde b_e/{\rm GeV}$ & $0.0726524$ & $\tilde c_e/{\rm GeV}$ & $1.23426$ \\ \hline
$s_{12}$ & 0.558474 &$s_{23}$ & 0.71674 &$s_{13}$ & 0.1497 \\ \hline
$\Delta m^2_{\rm sol}$ & $7.47522\times 10^{-5}~{\rm eV}^2$ &  $\Delta m^2_{\rm atm}$ & $2.53455\times 10^{-3}~{\rm eV}^2$ &
$\delta_{\rm CP}$ & $277.664^{\circ}$ \\ \hline
$\alpha_{21}$ & $0.442^{\circ}$ & $\alpha_{31}$ & $180^{\circ}$ & $\kappa$ & $76.3653~{\rm meV}$ \\ \hline
$m^2_{\nu e}$ & $0.023038~{\rm eV}^2$ &$\sum D_{\nu}$ & $447.588~{\rm meV}$ & $\langle m_{ee}\rangle$ 
& $151.663~ {\rm meV}$ \\
\hline
    \end{tabular}
    \caption{\label{tab:BF}%
      Best-fit (BF) parameter values in the IH case corresponding to $\Delta\chi_{\rm min} =2.14$, where $\tilde a_e\equiv \frac{vv'}2 \frac{a_e f_1}{M_1}$, $\tilde b_e\equiv \frac{vv'}2 \frac{b_e f_2}{M_2}$, and $\tilde c_e\equiv \frac{vv'}2 \frac{c_e f_3}{M_3}$.}
\end{table}

\section{Summary and discussion}
\label{sec:IV}
We have proposed a lepton seesaw model in a modular $A_4$ symmetry. 
The charged-lepton mass matrix and the neutrino mass matrix have the same origin associated with vector-like fermions.
As a result, the charged-lepton mass matrix is induced via seesaw-like mechanism, while the neutrino mass matrix is generated through the inverse seesaw mechanism where the tiny Majorana mass matrix is supposed to be obtained via VEV of $\Delta$.

In addition we have searched for allowed regions at nearby fixed points which are favored by a flux compactification of Type IIB string theory~\cite{Ishiguro:2020tmo}, and we have found the solutions at nearby $\tau=\omega$.
In our chi square numerical analysis, we have found large value of sum of the neutrino masses which exceed the cosmological upper bound $0.12$ eV for both the cases of NH and IH.
However, our solutions are totally safe for the direct search for neutrino mass which is done by the Karlsruhe Tritium Neutrino experiment for both the cases. 
Our region of neutrinoless double beta decay is localized at nearby the experimental bound, therefore it would be tested near future. We have obtained narrow ranges of Majorana and CP phases which would also be verifiable via some experiments.

Finally, we would like to mention our next task.
Even though we have successfully built our lepton seesaw model which has the same mass origin from vector-like fermions, 
our model has an assumption among mass hierarchies such as $m'\ll M_{L'}$.
This assumption is very important to construct the seesaw model buildings.
Thus, we will move to a model that would have appropriate and natural mass hierarchies among these mass matrices.

\section*{Acknowledgments}
The project is supported by the Fundamental Research Funds for the Central Universities (T.~N.) {
and Zhongyuan Talent (Talent Recruitment Series) Foreign Experts Project (H.~O.)}.

\bibliography{ctma4.bib}
\end{document}